\documentclass[12pt,preprint]{aastex}

\newcommand{\kms}{$\mathrm {km\,s}^{-1}$}
\newcommand{\mdot}{$\mathrm {M_{\odot}\,yr}^{-1}$}

\shorttitle{Vortices in the wakes of AGB stars}
\shortauthors{Wareing et al.}

\begin{document}

\title{Vortices in the wakes of AGB stars}

\author{C. J. Wareing\altaffilmark{1}, Albert A. Zijlstra\altaffilmark{1} and T. J. O'Brien\altaffilmark{2}}
\altaffiltext{1}{School of Physics and Astronomy, University of Manchester, Campus North, PO Box 88, Manchester, M60 1QD, UK; c.j.wareing@manchester.ac.uk, a.zijlstra@manchester.ac.uk}
\altaffiltext{2}{Jodrell Bank Observatory, University of Manchester, Macclesfield, SK11 9DL, UK; tim.obrien@manchester.ac.uk}

\begin{abstract}

Vortices have been postulated at a range of size scales in the universe including
at the stellar size-scale. Whilst hydrodynamically simulating the
wind from an asymptotic giant branch (AGB) star moving through and sweeping up its 
surrounding interstellar medium (ISM), we have found vortices on the size scale of $10^{-1}$ 
pc to $10^1$ pc in the wake of the star.
These vortices appear to be the result of instabilities at the head of
the bow shock formed upstream of the AGB star. The instabilities peel off downstream 
and form vortices in the tail of AGB material behind the bow shock, mixing with 
the surrounding ISM. We suggest such structures are visible in the planetary 
nebula Sh 2-188.

\end{abstract}


\section{Introduction}

Vortex instabilities are observed throughout the universe. On the smallest scales
in the laboratory, \cite{robey02} have experimentally investigated the 3D
interaction of a strong shock with a spherical density inhomogeneity via
laser-driven experiments. They found a double vortex ring structure and an 
azimuthal instability that ultimately results in the breakup of the 
ring. They also performed 3D computational
simulations which were shown to be in remarkable agreement with their experimental 
investigations. 

In planetary atmospheres, vortices can be regularly observed in von Karman vortex 
streets often associated with atmospheric flow past stationary objects such as 
mountains. They have also been suggested as the explanation of the giant storm 
on Jupiter known as the Great Red Spot \citep{williams97}. Atmospheric simulations
have successfully reproduced vortices comparable to the Great Red Spot 
\citep{yamazaki04} and other vortex anomalies in the atmosphere of Jupiter 
\citep{morales05}. \cite{li01} have hydrodynamically simulated 
vortices at the stellar size scale in thin accretion disks.


Vortices at the interstellar size scale have been predicted since some of the
first 'high' resolution calculations by \cite{woodward76} where Kelvin-Helmholtz
instabilities begin the breakup of interstellar clouds. More recently,
\cite{klein94} predicted the existence of such vortices forming from
the interaction of strong shocks with the medium through which they propagate.
A series of laboratory experiements as well as two- and three-dimensional 
simulations \citep{klein03} confirmed these earlier predictions.
They modelled the interaction of a supernova 
shock and a cloud in the interstellar medium (ISM) as 
the effect of a high density sphere embedded in 
a low density medium after the passage of a strong shock wave.
They found vortex rings are shed from the sphere and, importantly,
these vortex rings break up at high Mach number due to three-dimensional bending-mode 
vortex-ring instabilities. Since their simulations are isothermal, they 
can be scaled to any size, for the given Mach number, and so are
relevant to clouds in the ISM ranging from the smallest resolved
structures at $\sim10^{-3}$ pc to diffuse clouds up to 100 pc.

Vortices can act as a source
of local angular momentum and turbulence, both of which are important in the evolution
of molecular clouds. For a discussion of the importance of interstellar
turbulence, we refer to the two-part review by Elmegreen \& Scalo (2004a,b).
Such molecular clouds are themselves important for star 
formation \citep{tilley04}. Vortices in
the ISM can also affect anything expanding into this medium e.g. 
supernovae, novae and planetary nebulae.

In this paper, we present simulations that show under certain conditions 
von Karman-like vortices are produced in the wakes of asymptotic giant branch 
(AGB) stars with size scales from $10^{-1}$ pc to $10^1$ 
pc. Specifically, we model the interaction of a stellar wind with the ISM where 
the star has a significant motion through the ISM. The stellar wind in our simulation
is produced by an AGB star \citep{wareing06b}. 
AGB stellar winds have typical speeds of $\sim 15$ \kms\ and mass-loss rates 
$\dot M \sim10^{-7}$--$10^{-5}$ \mdot. This wind
ejects up to a few solar masses of material into the surrounding medium
over the course of $\sim 500\,000$ years.
In our two-wind model, we have found that the movement of the star through the ISM
(modelling the ISM as a 
second wind) causes the ejected stellar material to form into a bow
shock upstream of the star with a tail flowing downstream. It is in this tail
that we find evidence of von Karman-like vortices. Observational evidence for interaction
between the ISM and AGB winds has been found by \citet{zijlstra02} and \citet{schoier05} and a model
of the interaction has been used to explain the shape of the 
structure around R Hya as a bow shock \citep{ueta06,wareing06b}.

In hindsight, previous 2D astrophysical simulations considering
this phase of stellar evolution have produced such vortices 
\citep[see figure 18 of][]{szentgyorgyi03}, but the authors did not discuss them. 
We consider
the vortices in 3D allowing an investigation and description of their behaviour.

\section{Simulations}

A computational fluid dynamics (CFD) scheme, {\sc cubempi}, has been used to 
hydrodynamically simulate the interaction of a star moving through the ISM 
ejecting a stellar wind during the AGB phase of evolution. The numerical 
scheme is based on a second-order Godunov scheme due to \citet{falle91}. 
The Riemann solver is due to \citet{vanleer79}. The scheme is Eulerian, posed in
three dimensions and cartesian coordinates and is fully parallel using the
MPI\footnote{http://www-unix.mcs.anl.gov/mpi/} library of subroutines. 
The scheme also includes the effect of radiative 
cooling via a parametric fit to the cooling curve of \citet{raymond76} 
above $10^4$ K. The scheme has been fully tested using standard CFD tests 
and its performance in these tests compared to results obtained by \citet{liska03}
for other CFD schemes. Astrophysical tests have also been performed using the
Sedov and Primakoff analytical supernova models. The full details of the scheme
and its performance can be found in \citet{wareing05}. 

We have used a two-wind model with a numerical domain of $200^3$ cells. This model
is the same as that used to successfully model the structure around R Hya 
\citep{wareing06b}. It is also the
AGB stage of our triple-wind model following the AGB and post-AGB phases of
evolution
used to successfully model the planetary nebula (PN) Sh 2-188 \citep{wareing06}. 
We have considered four values of the mass-loss rate on the AGB: $10^{-7}, 
5\times10^{-7}, 10^{-6}\ \&\ 5\times10^{-6}$ \mdot. We have considered the AGB
phase, and hence the simulation, to last 500\,000 years with an AGB 
wind velocity of 15 \kms. These wind parameters are typical of previous models
in the literature \citep{mellema91,frank94}. An unphysical temperature of $10^4$ K 
was used for the AGB wind but this does not affect the overall result \citep{wareing06b}.
In view of the still considerable 
uncertainties on the detailed properties and evolution of these winds, more 
detailed temporal variations were not modelled. Relative velocities of 
the star with
respect to the ISM have been considered from 0 to 200 \kms\ in steps of 25 \kms\
corresponding to the range of velocities found within the Galaxy.
With speeds of relative movement above 75 \kms, we have not simulated mass-loss 
rates of $10^{-7}$ \mdot\ due to CPU constraints. The ISM has been modelled with
densities n$_{\rm H}$ = $2,\ 0.1\ \&\ 0.01\ \mathrm{cm}^{-3}$ and a temperature of $8\times10^3$ 
K. These are the properties of the warm ISM or WIM, the larger constituent of the
ISM \citep{burton88}. 

The simulations begin at the start of the AGB phase of evolution of the 
central star and are performed in the frame of reference of the star. 
The wind from the star drives a shock into the ISM. In the case of a
stationary star we have found the shock drives a spherical shell of 
AGB wind material sweeping up ISM material. Inside 
the reverse shock is a bubble of undisturbed AGB wind material. 
This bubble is of low density and temperature, increasing towards the star, 
and of high pressure. If the central star is moving through the ISM, 
a bow shock forms upstream of the star. Such
bow shocks have been simulated in other cases of winds interacting with injected
flows \citep{villaver03,pittard05}. Eventually,
the simulations show the bow shock reaches a maximum distance ahead of star 
which can be understood in terms of a ram pressure balance between the stellar 
wind and the ISM. Strong shock theory predicts the temperature of the shocked 
material at the head of the bow shock: T $\sim 3/16\ m\,v^2/k$ in general
agreement with the simulations. 
Ram-pressure-stripped material from the head of the
bow shock forms a tail behind the nebula. As material moves down the tail,
it cools and mixes with ISM material. 

\section{Results}

\begin{figure}
\begin{center}
\epsscale{0.5}
\plotone{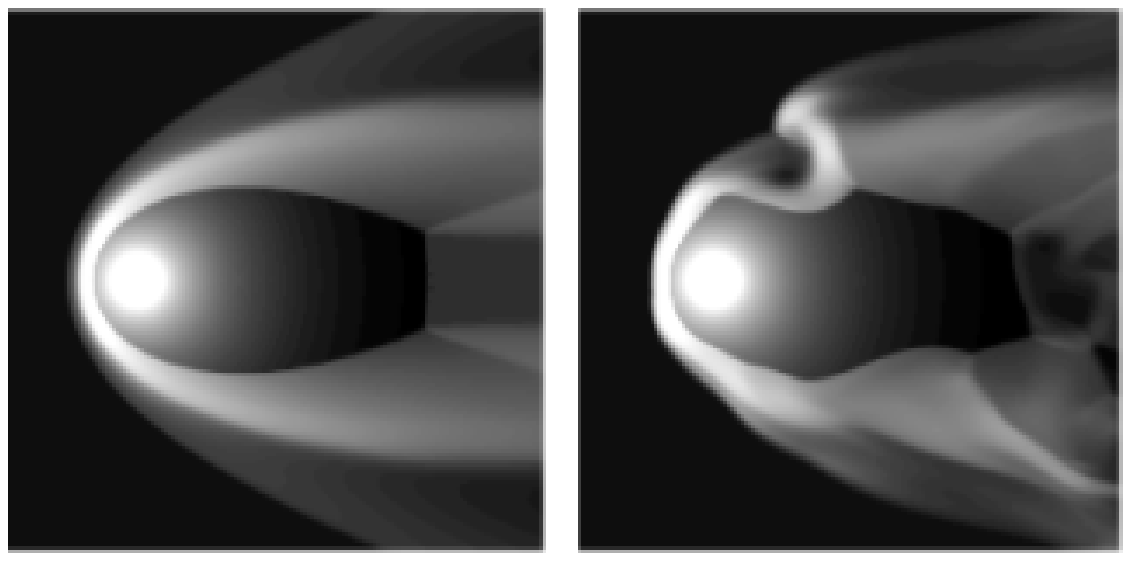}
\caption{Snapshots of the logarithm of density from simulations 
of the AGB evolution of a central star moving at 75 \kms. The ISM 
is flowing from left to right. Both simulations have an ISM density 
n$_{\rm H}$ = 2 cm$^{-3}$. The left panel has a mass-loss rate of 10$^{-7}$ \mdot\ 
and is 0.25 pc on a side. The right panel has a mass-loss rate of
$5 \times 10^{-6}$ \mdot\ and is 1.75 pc on a side.}
\label{density}
\end{center}
\end{figure}

Figure \ref{density} shows two snapshots from the AGB evolution of a central 
star moving at 75 \kms. Both panels show the gas state at the end of the AGB phase,
500\,000 years into the simulation. In the left panel, the bow shock driven 
by the AGB wind has reached a stationary position at the point of ram 
pressure balance, 0.02 pc ahead of the central star, with a smooth tail of 
shocked material extending downstream of the bow shock. In the right panel, 
where the mass-loss rate is 50 times higher, the tail contains vortices 
flowing downstream.

\begin{figure}
\begin{center}
\epsscale{0.5}
\plotone{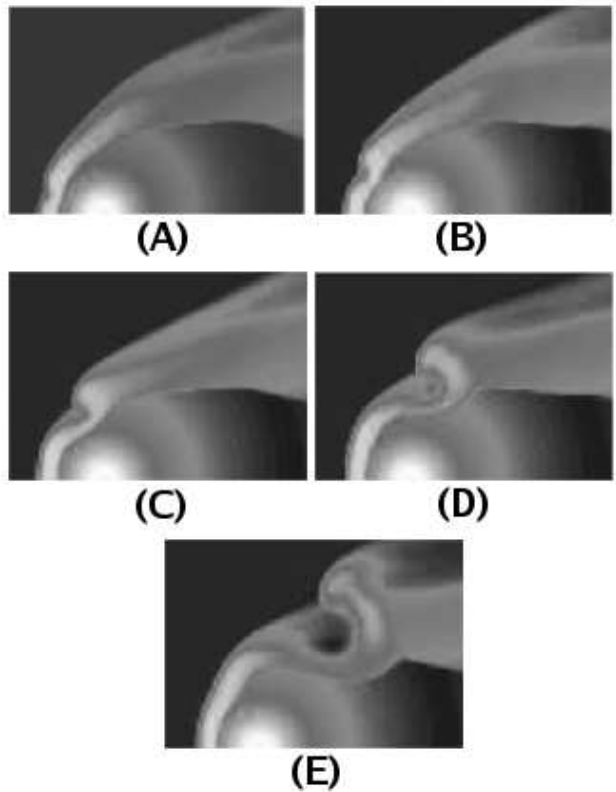}\\
\caption{Shedding of the vortex in detail. 
The panels show portions of slices of the logarithm of density through 
the position of the central star parallel to the direction of motion. 
Panel A is at 460\,000 years into the AGB phase, panel B 470\,000 years, 
panel C 480\,000 years, panel D 490\,000 years and panel E 500\,000 years.}
\label{evolution}
\end{center}
\end{figure}

Figure \ref{evolution} shows the evolution of the vortex in the right panel of
Figure \ref{density}. In the top panel, the indentation near the head of bow 
shock indicates an instability, 
which in the second panel is shed downstream from the head of the bow shock. 
The remaining panels 
show the instability forming into a vortex, spiralling and flowing downstream.
This episode of vortex shedding from the head of
the bow shock is not the first such event.
After 50\,000 years of evolution, the tail appears to 
become turbulent and the first episode of vortex shedding begins at 90\,000 years.
The initial episode appears to be symmetric around the bow shock and the instability
forms into a vortex ring flowing downstream.
The next episode begins at 205\,000 years and continues with multiple
episodes of vortex shedding with no clear periodicity or ring symmetry.
The vortices flow out of the simulation domain in 100\,000 years having
achieved a stable structure which suggests they will have an extended lifetime
in the wake.

\begin{figure}
\begin{center}
\epsscale{0.5}
\plotone{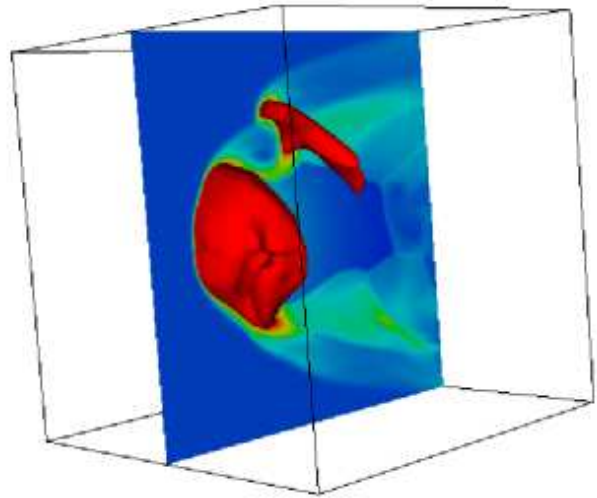}
\caption{Visualisation of the simulation datacube showing gas density using 
the visualisation software package {\sc mayavi} at the same time as Figure \ref{density}. 
A planar slice through the position of the central star and parallel to the direction
of motion is shown. This 
slice is intersected by a 3D isosurface of constant density (in red) demonstrating the 
shape of the bow shock with a vortex arc further downstream. Note that the vortex
downstream has become azimuthally unstable.}
\label{3d}
\end{center}
\end{figure}

To show the breakdown of the later vortices azimuthally, Figure \ref{3d} shows a 
visualisation of the density datacube 500\,000 years into the simulation.
An azimuthally-unstable partial vortex ring can be seen mid-way down the tail 
and more partial vortex-rings are in the process of shedding off the bow shock.

In order to investigate the effect of grid resolution on the production of vortices,
the same simulation which produces vortices with a movement through the ISM of 75 \kms\
was rerun at a grid resolution of $100^3$ and $300^3$. 
The low resolution simulation showed a smooth, steady state tail with no clear 
vortices produced from the head of the bow shock. Undulations at the head of the bow 
shock do appear to move down the tails. We suggest the initial instabilities 
which lead to vortices downstream are strongly suppressed by this low resolution.
At $300^3$, many more vortices than at $200^3$ seem to be produced from similar 
instabilities at the head of the bow shock, but
importantly the physical size of the vortices is similar and therefore not dependent on resolution.
A 2D numerical simulation was also performed in cylindrical polar coordinates
for further comparison. Similar vortex-like structures 
form in the tail of AGB material and flow downstream. The initial instability
appears to be slower in developing and once developed, the 
instability causes a periodic movement of the tails more like
von Karman vortices. The vortices themselves appear as density enhancements
moving down the tail structure. Clearly in 3D simulations the vortices 
form fully and the periodicity is destroyed by the instability of
the bow shock following the launching of the first vortex ring.

\section{Discussion}

We postulate that the vortices in the tail are the result of vortex shedding 
from the head of the bow shock. Vortex shedding occurs when a fluid passes by 
an object and the shear layer near the object creates
a velocity gradient. The Kelvin-Helmholtz instability often accompanies flows 
where shear is present, making the flow unstable to perturbations.
The simulations show that the bow shock does not remain 
stationary in the frame of reference of the star after it has reached the 
ram pressure balance point. In fact, its distance ahead of the star is 
oscillating and it is the Kelvin-Helmholtz instability which
is responsible for this oscillation. As the instabilities spread from the head of the 
bow shock, peel off and flow downstream to form vortices, the bow shock loses material
in the vortices and is driven back towards the star by the pressure of the ISM.
As the stellar wind supplies more material to the bow shock, it recovers its
ram-pressure-balance position and this instability-driven oscillation begins 
again. Note that this 
shows that instabilities set in only after the bow shock has reached its balance
position, and thus vortices only appear in the tail after this time.

\begin{figure}
\begin{center}
\epsscale{0.5}
\plotone{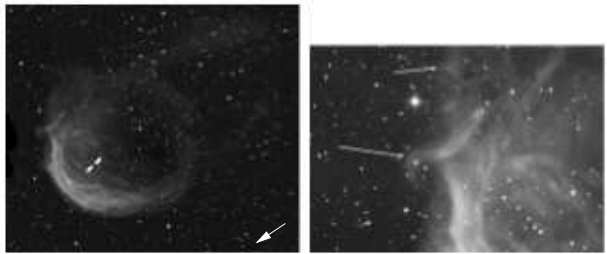}
\caption{H$\alpha$ images showing the PN Sh 2-188. On the left is a combined image
of the whole nebula showing the bright south-eastern arc and the faint north-western
structures interpreted as a bow shock and tail respectively. The nebula is 11 arcminutes
in diameter with a tail of 18 arcminutes. In the left panel, the central star is indicated
between the markers, with the arrow indicating its proper motion. For
more information, refer to \protect\cite{wareing06}. We interpret the structures 
indicated by arrows in the right panel as instabilities being shed from the bow-shock.
The images are mosaics taken as part of the Isaac Newton Telescope Photometric
H$\alpha$ Survey of the Northern Galactic Plane (IPHAS) \citep{drew05} in 2003. North
is up and East is to the left. The left image is $20 \times 16.7$ arcmin. and the 
right image is $7.6 \times 5.3$ arcmin.}
\label{sh2-188}
\end{center}
\end{figure}

Figure \ref{sh2-188} shows the PN Sh 2-188 in the left panel and a detail of the
nebula in the right panel. In a previous paper, \cite{wareing06} modelled
Sh 2-188 as a nebula shaped by wind-ISM interaction where the bright arc
is a bow shock ahead of the central star. The structures highlighted
in the right panel can be interpreted in our vortex-shedding scenario as 
instabilities being peeled off the bow-shock which will form vortices
downstream.

Theoretically, vortices form over a certain range of Reynolds number. The Reynolds 
number of a flow is defined as the ratio of internal forces to viscous forces and 
can be formulated as $Re = \rho\ V\ D / \nu$ where $\rho$ is the fluid density,
$V$ is the free-stream fluid velocity, $D$ is the characteristic length and $\nu$
is the dynamic fluid viscosity. For a 
Reynolds number of less than 50, the wake behind an object 
is thin and the flow can be classed as laminar. With a Reynolds number 
around 100, features similar to the
von Karman vortex street develop as we see here. The wake also can become several 
times wider than the characteristic width of the object.
At Reynolds numbers of 1000 and greater, the wake behind 
the object becomes turbulent. It is difficult to estimate the Reynolds numbers in
our simulations as this involves the viscosity of the fluid and the simulations include an
artificial viscosity to suppress the Quirk instability \citep{quirk94}. Since this artificial
viscosity is constant in all the simulations, it is possible to estimate {\it relative}
Reynolds numbers between the simulations. The characteristic length in a particular 
simulation is estimated as the diameter of the object in the flow, defined by 
twice the ram-pressure-balance distance. Note that the point of ram pressure balance 
is inversely proportional to flow-velocity and thus so is the characteristic length.
Therefore, there is no dependency on flow-velocity in the estimation of relative Reynolds
numbers. The relative Reynolds number increases with increasing mass-loss rate and
increasing ISM density. In the simulations, there is a general trend to see 
vortices when the mass-loss rates or ISM densities are not extreme. These simulations
correspond to the mid-range of relative Reynolds numbers. However, in some cases of extreme
mass-loss rate or ISM density, the bow shock structure is either so large that it is possible we
do not see vortices on the simulation timescale, or so small that it is possible the ISM
flow confines the bow shock enough to suppress the formation of vortices.
Thus, it is not clear
whether the vortices are dependent on relative Reynolds number although it seems
there is a range where vortices are more apparent on the
timescales of these AGB simulations, which we might expect from theoretical indications. All the 
simulations are supersonic from a Mach number of 1.8 at 25 \kms\ to 14.4 at 200
\kms\ suggesting no dependency on Mach number.

The vortex shedding lacks the regular period seen in the 
case of von Karman vortices. We do note though that the first episode of vortex shedding 
is in the form of a symmetric ring shed downstream from the head of the bow shock. Further
episodes of vortex shedding do not shed such stable rings and vortices are shed from either
side of the bow shock. This can be understood in terms of the regular shape of the bow shock at
the time of the first vortex shedding instability; following this, vortex shedding occurs
from an irregular bow shock and results in irregular behaviour.

AGB winds are important sources of dust and light elements
and thus for the evolution of the global ISM, but their importance to the
dynamics of the ISM has not previously been considered. Our results show that
several vortices can be launched into the ISM during the AGB phase of
evolution. These improve the mixing of material, and provide the local ISM
with a source of a angular momentum. Our simulations suggest the vortices are
long-lived since they appear to form stable structures which flow out of the
simulation domain. 

The space density of mass-losing AGB stars is low ($\sim
100$ kpc$^{-3}$), and assuming a typical life time of the phase of
$10^5$ yrs, each region in the ISM may be affected by the tail of an AGB
star every $10^9$ yrs. We estimate the power input into the ISM from AGB
winds to be $2.5 - 125 \times 10^{-32}$ erg\,cm$^{-3}$\,s$^{-1}$. This is
several orders of magnitude less than the contribution of supernovae and
winds from massive stars at around $10^{-25}$ erg\,cm$^{-3}$\,s$^{-1}$
\citep{elmegreen04a}. We find the temperature of material in the vortices 
to be on the order of 5-10\,000 K, cooling as the vortex moves downstream.
The global (as opposed to the local) importance of the vortices may therefore 
be limited and possibly the angular momentum returned to the local ISM may be 
most important. It is also possible that the galactic distribution of AGB stars
provides a source of turbulence where supernovae and winds from massive 
stars do not. One may also expect similar
vortices to form behind other types of mass-losing stars. We suggest that the 
swirls seen in the light echo of V838 Mon \citep{bond06} could represent such 
vortices.

In our simulations we have modelled the ISM with a constant density. In reality,
the ISM is unlikely to be homogeneous and this will more than likely
affect the structures forming around the star. It is possible that inhomogenities
would fracture the bow shock, as seen in the case of the PN Sh 2-188 \citep{wareing06},
and seed more vortex shedding. Further, a star travelling at 200 \kms\ would travel
approximately 100 pc during the AGB phase and with a bow shock cross-section of up
to $\sim 2.0$ pc$^{\rm 2}$, interact with a volume 
of ISM of up to 200 pc$^3$, which could contain on the order of a few tens of stars
which could interact with the vortices.

In our simulations, we have neglected viscosity and observed a 
variety of behaviour from low to high relative Reynolds numbers. Magnetic fields may 
suppress viscosity significantly implying a greater degree of turbulence. However, 
large amounts of viscosity in the ISM would smooth the structures which develop in 
our simulations. It is for this reason that it would be premature to quantify
when to expect vortices for particular stellar parameters. 

\acknowledgments

The authors acknowledge very useful comments from the anonymous referee and Professor
John Scalo. CJW acknowledges the support of PPARC.
The numerical computations were carried out using the COBRA supercomputer at Jodrell
Bank Observatory.

\clearpage

\clearpage

\clearpage
 
\clearpage
 
\end{document}